\documentclass[twocolumn,prl,groupedaddress]{revtex4}%
\usepackage{amssymb}
\usepackage[dvips]{graphicx}
\usepackage{dcolumn}
\usepackage{bm}
\usepackage{amsmath}
\usepackage{amsfonts}%
\begin{document}
\title{Prediction of the capacitance lineshape in two-channel quantum dots}
\author{C.~J.~Bolech}
\affiliation{Universit\'{e} de Gen\`{e}ve, DPMC, 24 Quai Ernest Ansermet, CH-1211
Gen\`{e}ve 4, Switzerland}
\author{N.~Shah}
\affiliation{Festk\"{o}rpertheorie, Paul Scherrer Institut, CH-5232 Villigen PSI \emph{and}
Theoretische Physik, ETH-H\"{o}nggerberg, CH-8093 Z\"{u}rich, Switzerland}
\altaffiliation{Present address: Institut f\"ur Theoretische Physik,}

\altaffiliation{Universit\"at zu K\"oln, 50937 K\"oln, Germany.}

\date{June 1$^{\text{st}}$, 2004}

\begin{abstract}
We propose a set-up to realize two-channel Kondo physics using quantum dots.
We discuss how the charge fluctuations on a small dot can be accessed by using
a system of two single electron transistors arranged in parallel. We derive a
microscopic Hamiltonian description of the set-up that allows us to make
connection with the two-channel Anderson model (of extended use in the context
of heavy-Fermion systems) and in turn make detailed predictions for the
differential capacitance of the dot. We find that its lineshape, which we
determined precisely, shows a robust behavior that should be experimentally verifiable.

\end{abstract}
\maketitle

Forty years have lapsed since Kondo published his celebrated article
\cite{kondo1964}. Even today, the Kondo effect continues to be one of the
central themes in condensed matter physics. To add to its importance in the
context of heavy fermions and localized magnetic moments in metals, one should
include now a growing set of experiments that during the last years observed
it in quantum dots and other mesoscopic devices
\cite{cronenwett1998,goldhaber1998}. The great appeal of such experiments
resides in their high degree of control over the different parameters involved
that yields unprecedentedly accurate and comprehensive tests for the theories.

An important variant of the Kondo problem is its multi-channel generalization
that is known to give rise to non-trivial low-temperature physics
characterized by fractional exponents and divergencies \cite{ad1984}. Mainly
due to its sensitivity to magnetic fields and channel anisotropy, the
experimental examples of multi-channel Kondo physics remain largely
controversial. Partly for this reason, as well as for the inherent richness of
the physics involved, a large amount of recent theoretical
\cite{matveev1991,matveev1995,gramespacher2000,oreg2003,lebanon2003a,lebanon2003b,shah2003,pustilnik2004,florens2004,anders2004a}
and experimental \cite{ralph1994,berman1999} efforts seek to realize two- or
multi-channel Kondo systems in the controlled realm of artificial
semiconducting nanostructures and other mesoscopic devices. In this letter we
want to contribute to that effort by proposing a highly tunable quantum-dot
system that allows access to the two-channel charge-fluctuation physics which
lies beyond the scope of an effective Kondo Hamiltonian description.

\begin{figure}
[t]
\begin{center}
\includegraphics[
height=1.4719in,
width=3.2007in
]%
{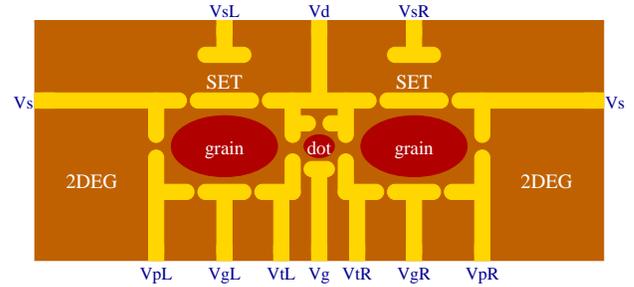}%
\caption{Cartoon micrograph of the proposed set-up: a small quantum dot (dot)
is connected to two larger dots (grains) that once the tuning is done are
isolated from the two-dimensional electron gases (2DEGs) at right and left. On
the top defining gates of the dot a system of two single electron transistors
(SETs) is built in. A set of gates controls the occupancy of dots and grains
($V_{g}$ and $V_{gL,R}$), their inter-coupling ($V_{t_{L,R}}$), and their
coupling to the 2DGEs ($V_{pL,R}$).}%
\label{fig: setup}%
\end{center}
\end{figure}

We propose a set-up involving one quantum dot linked to two mesoscopic islands
that we will refer to as \textit{grains}. The grains are weakly Coulomb
blockaded as compared to the dot and have a quasi-continuous energy spectrum
as against the discrete one of the dot. In Fig.~\ref{fig: setup} we illustrate
the configuration by\textbf{\ }showing the relative disposition of dot and
grains and by indicating the different gates. A set of voltage gates tunes the
occupancy of the dot ($V_{g}$) and the grains ($V_{gL}$, $V_{gR}$) while
another set ($V_{t_{L}}$, $V_{t_{R}}$) controls the couplings between dot and
grains. Although for the most part we envision the grains to be completely
decoupled from the two-dimensional electron gas (2DEG) on either side, the
indicated gate voltages $V_{pL}$ and $V_{pR}$ allow for controlling the
coupling during the tuning stage. Also, for comparison to the conventional
set-up of a dot coupled to two leads (that features single-channel physics),
the 2DEGs can be allowed to flood the grains by not energizing these two gates.%

The purpose of this set-up is to allow controlled experimental measurements of
the charging lineshape of the quantum dot for a wide range of $V_{g}$. Charge
fluctuations on the dot are due to the dot-environment coupling which in turn
is at the root of complex many-body behaviors such as the single- or
multi-channel Kondo effect. In our case the \textit{environment} is given by
the two grains when the plunger voltages ($V_{pL}$, $V_{pR}$) are completely
pinched off in order to decouple the dot-grains system from the 2DEGs. This
special environment was carefully engineered to gain a robust access into the
rather elusive regime of two-channel Kondo physics and mixed valence.
On\textbf{\ }one hand the isolated nature of the set-up forbids charge
transport experiments, but on the other hand it leaves open the possibility of
capacitance lineshape measurements that provide the most accurate and direct
information about charge fluctuations and dot-environment interplay.

For making these kind of measurements, the set-up should be prepared as
follows. First, let the two 2DEGs flood the grains while energizing
$V_{t_{\alpha}}$ so that the charge of the dot is quantized; apply a small
bias, and tune $V_{g}$ while measuring the current until the system is in the
middle of a Coulomb peak (a charge degeneracy point of the dot). Second,
energize $V_{p\alpha}$ but without isolating the grains and keep adjusting the
three voltage gates ($V_{g},$ $V_{gL},$ $V_{gR}$) in order to remain at the
center of the Coulomb peak. Slowly increase the values of the plunger gates
while keeping on adjusting as above, until no more current flows and the
dot-grains system is isolated from the 2DEGs. This way the dot as well as the
grains are all independently tuned near their charge degeneracy points.

As demonstrated in Ref.~%
[\onlinecite{berman1997}]%
, a high sensitivity charge sensor can be implemented using a metallic
Single-Electron Transistor (SET). In our proposed set-up, a \textquotedblleft
twin SET\textquotedblright\ (consisting of two SETs arranged in parallel and
each associated to one of the two grains) is incorporated directly into the
defining gates of the dot-grains system. The central island of each SET is
coupled to the source and drain electrodes ($V_{s}$ and $V_{d}$) via two small
tunnel junctions that make for the leading contribution to the SET capacitance
and must be kept as small as possible. One extra gate for each SET ($V_{sL}$
and $V_{sR}$) permits independent fine tuning to achieve the regime of maximum
charge sensitivity.

We now turn to the task of writing down a Hamiltonian for this set-up. The
largest energy scale involved is the dot charging energy ($E_{d}$) followed by
the charging energies of the grains ($E_{g\alpha}$, $\alpha=L,R$), each
responsible for the respective Coulomb blockade. The first step in the
modeling should then be to write down what we call the configurational energy
of the dot-grains system:
\[
E_{\mathrm{conf}}\!=\!E_{d}\left[  \hat{n}_{d}-\left(  N_{d}+y\right)
\right]  ^{2}+\sum\nolimits_{\alpha}\!\!\!E_{g\alpha}\left[  \hat{n}_{g\alpha
}\!-\left(  N_{g\alpha}\!+x_{\alpha}\right)  \right]  ^{2}%
\]
Here we have assumed for the moment that the dot and the grains are all
individually isolated and that their respective\ energies change as the value
of the voltage in the corresponding\ gates is varied (with the
linear-function\ identifications\textbf{\ }$V_{g}\mapsto y$\textbf{\ }%
and\textbf{\ }$V_{g\alpha}\mapsto x_{\alpha}$). The notations $\hat{n}_{d}%
$\ and $\hat{n}_{g\alpha}$\ correspond to the number operators, while $N_{d}%
$\ and $N_{g\alpha}$\ to certain ground state occupations\ for\textbf{\ }%
$y=x_{\alpha}=0$, of the dot and the grains. Right after preparing the system
the way it was described above, we have that $y\approx x_{\alpha}\approx1/2$
and that $N_{d}$ and $N_{g\alpha}$ are fixed to certain undefined integer
values (we will restrict ourselves to variations\ of\textbf{\ }$y$%
\textbf{\ }and $x_{\alpha}$\ in the interval $\left[  0,1\right]  $). We are
going to consider the dot-grains system in isolation; this amounts to imposing
the constraint of total charge conservation, that when exactly one
\textit{extra} electron is captured reads:
\begin{equation}
\hat{n}\equiv\hat{n}_{d}+\sum\nolimits_{\alpha}\hat{n}_{g\alpha}=\left(
N_{d}+1\right)  +\sum\nolimits_{\alpha}N_{g\alpha}\equiv N+1
\label{eqn: constraint}%
\end{equation}

In what follows, we are going to consider $x_{\alpha}$\ to be fixed and let
only $y$\ vary. The charge of the dot fluctuates between $e(N_{d}%
+1)\leftrightarrow eN_{d}$. In the slave-operator language, we associate a
fermion operator $\hat{f}_{\sigma}^{\dagger}$ to the configuration when the
extra electron (with spin $\sigma$) is on the dot and a boson operator
$\hat{b}_{\bar{\alpha}}^{\dagger}$ to the configuration when the dot is
`empty' and the electron is on the grain $\alpha$. The boson does the job of
book-keeping by indicating that back tunneling into the dot coming from the
opposite grain is blocked ($\bar{\alpha}$\ stands for the opposite of $\alpha
$). The respective configurational energies read
\begin{align}
\varepsilon_{f}  &  =E_{d}\left(  1-y\right)  ^{2}+\sum\nolimits_{\alpha
}E_{g\alpha}\left(  x_{\alpha}\right)  ^{2}\\
\varepsilon_{b\alpha}  &  =E_{d}\left(  y\right)  ^{2}+E_{g\alpha}\left(
1-x_{\alpha}\right)  ^{2}+E_{g\bar{\alpha}}\left(  x_{\bar{\alpha}}\right)
^{2}%
\end{align}
We will make the simplifying assumption that $\varepsilon_{bL}=\varepsilon
_{bR}\mapsto\varepsilon_{b}$; this is true, for instance, in the case of a
symmetric set-up (the implications of relaxing this symmetry will be discussed
later). The reader can convince himself that all other charge configurations
are either forbidden due to the constraint of Eq.~(\ref{eqn: constraint}) or
are much higher in energy and can be safely ignored. We will concentrate on
the case of $N_{d}$ even, since, as we shall see, the symmetry of the model is
$SU\!\left(  2\right)  \otimes SU\!\left(  2\right)  $ and displays
two-channel Kondo physics of interest to us. The case of odd $N_{d}$ is also
interesting and\ will be discussed elsewhere (it gives rise to a
single-channel $SU\!\left(  4\right)  $ model; cf.~Ref.~
[\onlinecite{lehur2004}]%
).

We shall model the quasi-continuum spectra ($\varepsilon_{k\sigma}^{\alpha}$)
of the grains by a flat density of states ($\rho_{g\alpha}$) and denote the
electron creation operator in the grain by $\hat{g}_{k\alpha\sigma}^{\dagger}%
$. Writing the configurational energy in Hamiltonian language and adding a
term describing the tunneling between dot and grains, we have the total
Hamiltonian
\begin{align}
\hat{H} &  =\sum\nolimits_{\alpha}\hat{H}_{\mathrm{grain}}^{\alpha}+\hat
{H}_{\mathrm{conf}}+\hat{H}_{\mathrm{tun}}\label{eqn: hamil}\\
\hat{H}_{\mathrm{grain}}^{\alpha} &  =\sum\nolimits_{k\sigma}\varepsilon
_{k\sigma}^{\alpha}\,\hat{g}_{k\alpha\sigma}^{\dagger}\hat{g}_{k\alpha\sigma
}^{%
\phantom{\dagger}%
}\\
\hat{H}_{\mathrm{conf}} &  =\sum\nolimits_{\sigma}\varepsilon_{f}\,\hat
{f}_{\sigma}^{\dagger}\hat{f}_{\sigma}^{%
\phantom{\dagger}%
}+\sum\nolimits_{\alpha}\varepsilon_{b\alpha}\,\hat{b}_{\bar{\alpha}}%
^{\dagger}\hat{b}_{\bar{\alpha}}^{%
\phantom{\dagger}%
}\\
\hat{H}_{\mathrm{tun}} &  =\sum\nolimits_{k\alpha\sigma}t_{k\alpha}\left[
\hat{g}_{k\alpha\sigma}^{\dagger}\hat{b}_{\bar{\alpha}}^{\dagger}\hat
{f}_{\sigma}^{%
\phantom{\dagger}%
}+\hat{f}_{\sigma}^{\dagger}\hat{b}_{\bar{\alpha}}^{%
\phantom{\dagger}%
}\hat{g}_{k\alpha\sigma}^{%
\phantom{\dagger}%
}\right]
\end{align}
The last two terms in the Hamiltonian together with the Hilbert space
constraint (cf.~Eq.~(\ref{eqn: constraint}))
\begin{equation}
\sum\nolimits_{\sigma}\hat{f}_{\sigma}^{\dagger}\hat{f}_{\sigma}^{%
\phantom{\dagger}%
}+\sum\nolimits_{\alpha}\hat{b}_{\bar{\alpha}}^{\dagger}\hat{b}_{\bar{\alpha}%
}^{%
\phantom{\dagger}%
}=1
\end{equation}
encode the physics of the Coulomb blockade. As before, we make the simplifying
assumption $\varepsilon_{k\sigma}^{L}=\varepsilon_{k\sigma}^{R}\mapsto
\varepsilon_{k\sigma}$\ and $t_{kL}=t_{k^{\prime}R}\mapsto t$. We also define
$\Delta_{\alpha}\equiv\pi\rho_{g\alpha}t^{2}\mapsto\Delta$ and $\varepsilon
_{\alpha}\equiv(\varepsilon_{f}-\varepsilon_{b\alpha})\mapsto\varepsilon$.

Remarkably, the Hamiltonian $\hat{H}$ turns out to be the same as the one
originally studied by Cox in the context of uranium heavy fermions
\cite{cox1987}. Substantial progress was made on the analysis of this
two-channel Anderson model during recent years
\cite{schiller1998,koga1999,bolech2002,johannesson2003,anders2004b}. The
emerging picture indicates that the low energy physics is governed by a line
of boundary conformal-invariant fix points of two-channel non-Fermi-liquid
nature. Several thermodynamic properties display low-$T$ logarithmic
divergencies and transport properties show $T$-dependencies that deviate from
the Fermi-liquid laws of the single-channel models and extend into the
mixed-valence regime. In particular, detailed predictions can be made for the
charge susceptibility of the dot which, in contrast with the field
susceptibilities, displays a regular behavior \cite{bolech2002}.%

\begin{figure}
[t]
\begin{center}
\includegraphics[
height=2.3333in,
width=3.3278in
]%
{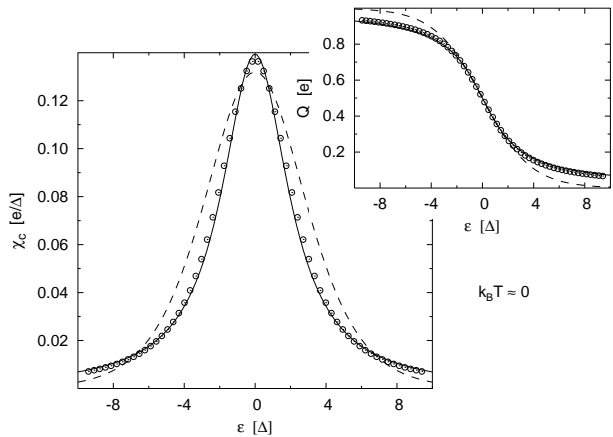}%
\caption{Charge susceptibility of the quantum dot as a function of energy
splitting. The open circles indicate the vanishing-temperature TBA result. The
solid line corresponds to a Lorentzian fit, while the dashed line is the fit
to the derivative of a Fermi-Dirac distribution. With the same symbols as
before, the overlay shows the excess charge on the dot as a function of energy
splitting.}%
\label{fig: xc_vs_eps}%
\end{center}
\end{figure}

Let us define the dot excess charge as $Q\!\equiv e\!\sum\nolimits_{\sigma
}\!\!\!<\!\!\hat{f}_{\sigma}^{\dagger}\hat{f}_{\sigma}^{%
\phantom{\dagger}%
}\!\!>$. Its variations caused by the gate voltage determine the differential
capacitance of the dot, $\delta C\varpropto\partial Q/\partial V_{g}$, that
can be measured using the twin SET set-up. Since $V_{g}\sim y\sim
\varepsilon\equiv\varepsilon_{f}-\varepsilon_{g},$\ we have $\delta
C\varpropto\chi_{c}\equiv-\,\partial Q/\partial\varepsilon,$\ which is nothing
but the charge susceptibility of the impurity in the two-channel Anderson
model. Using Thermodynamic Bethe Ansatz (TBA) one can in fact find an
expression for $Q\left(  \varepsilon\right)  $ at zero temperature,
\[
Q\left(  \varepsilon\right)  =\int_{-\infty}^{+\infty}\int_{-\infty}^{0}%
\frac{\left(  2z/\pi\right)  ~dz}{\cosh\left[  \frac{\pi}{2\Delta}\left(
x-z\right)  \right]  }\frac{\left(  x-\varepsilon\right)  ~dx}{\left[  \left(
x-\varepsilon\right)  ^{2}+4\Delta^{2}\right]  ^{2}}%
\]
and from it compute $\chi_{c}\left(  \varepsilon\right)  $. Alternatively, one
can extract these two quantities at an arbitrary finite temperature from the
numerical solution of the TBA equations.%

\begin{figure}
[t]
\begin{center}
\includegraphics[
height=2.3333in,
width=3.3278in
]%
{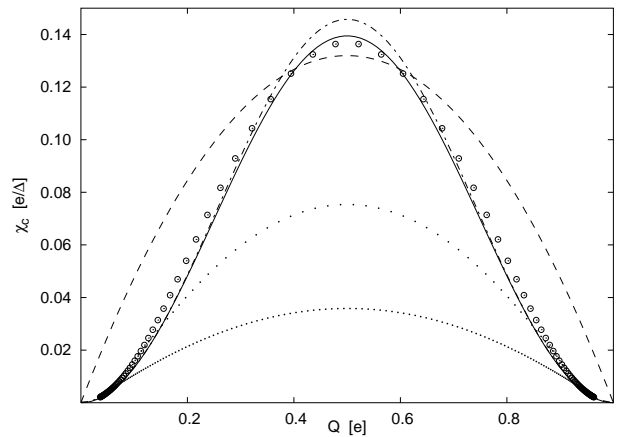}%
\caption{Low-temperature charge susceptibility of the quantum dot as a
function of its excess charge. The open circles are the results of the TBA
analysis and the solid and dashed lines are the translation of the Lorentzian
and Fermi-Dirac fit in the previous figure. The dash-dotted line is a fit to
the functional form appropriate for a different set-up with a single-grain.
The dotted lines are finite-temperature TBA results for two different
$k_{\mathrm{B}}T\gg\Delta$.}%
\label{fig: xc_vs_nc}%
\end{center}
\end{figure}

In Fig.~\ref{fig: xc_vs_eps} the results for the vanishing-temperature charge
susceptibility are given as a function of $\varepsilon$ (open circles). Two
commonly used fits are also plotted for illustrative purposes. The dashed line
corresponds to a fit to a \textquotedblleft purely thermally broadened
resonance\textquotedblright\ that was used successfully to fit the
very-weak-coupling regime of a single-grain set-up \cite{berman1999}. The
solid line, on the other hand, is a fit to a simple Lorentzian bell-shape. In
the figure overlay we show the excess charge as a function of $\varepsilon$,
together with the translation of the two fits in the main plot. In
Fig.~\ref{fig: xc_vs_nc} we present the plots of the charge susceptibility
(and the fits) as a parametrically defined (via $\varepsilon$) function of $Q$.

Our result can be characterized by two key features. First, one can observe
that $\chi_{c}$ is symmetric about $Q=0.5$ or $\varepsilon=0$. This symmetry
is a rather special property of the two-channel $SU\!\left(  2\right)  $
Anderson model, seen to be violated in other cases we shall touch upon later.
Second, $\chi_{c}$ is an universal function of $\Delta$, independent of the
quantitative details of the set-up tuning. This is a manifestation of having
only one scale in the mixed valence regime; below this scale the physics is
governed by the fixed point behavior \cite{bolech2002}. The differential
conductance lineshape turns out to be insensitive to temperature for a wide
range of low temperatures. It only starts to change appreciably for
$k_{\mathrm{B}}T\gtrsim\Delta$, when it eventually starts to become ever
flatter and quickly approaches the inverted-parabola shape characteristic of
thermal broadening; two such curves are shown by the dotted lines in
Fig.~\ref{fig: xc_vs_nc}. The story is similar as regards to the effect of an
external magnetic field. For low fields the lineshape remains unchanged until
$\mu_{\mathrm{B}}H\gtrsim\Delta$. Equivalent to the effect of an applied field
is the one of asymmetries between the tuning and characteristics of the right
and left grains. Only when those grain dissimilarities translate into
differences between $\varepsilon_{gL}$ and $\varepsilon_{gR}$ of the order of
$\Delta$ or larger, then deviations of the capacitance lineshape from the
predicted one are expected. If one were to use the Lorentzian form as a first
approximation to fit our exact prediction, it would mean that as long as the
asymmetries between the two grains and the temperature are well within
$\Delta$, one essentially has no free parameter after choosing the scale. This
is equivalent to fixing the area $A$ under the curve in the functional form
for the Lorentzian,
\begin{equation}
\chi_{c}\left(  Q\right)  =\frac{2A}{1+\tan^{2}\left[  \pi\left(
Q-1/2\right)  \right]  }\label{eqn: lorentz}%
\end{equation}

The symmetry of the line-shape which requires a delicate balance between the
number of channels and the internal degrees of freedom, combined with the
robustness against the effect of temperature and the details of tuning, is
what makes our prediction characteristically and clearly distinct for
experimental identification. One would like to tune $\Delta$ (via $V_{t_{a}}$)
to make it as large as possible. This will help also diminish the relative
magnitude of channel-hybridization asymmetries. Besides, $\Delta$ should be
bigger than the level spacing on the quasi-continuum of states in the grains.
The limitation is $\Delta\ll E_{d},E_{g\alpha}$ which should be the largest
energy scales in the problem. These practical constraints are not too
different from the ones for the single-grain set-up of Ref.~%
[\onlinecite{berman1999}]%
, what takes us to believe that the engineering challenges should be
surmountable. For the purpose of comparison, in Fig.~\ref{fig: xc_vs_nc} we
have displayed the fit (dash-dotted line) corresponding to the functional form
predicted (and experimentally verified) for the large-conductance regime of
the single-grain set-up consisting of a grain connected to a lead (see Refs.~%
[\onlinecite{matveev1995,berman1999}]%
). The related theory was based on mapping the charge susceptibility to the
spin susceptibility of an effective two-channel Kondo model and was thereby
expected to have a logaritmic divergence (required to be cut-off in matching
with the experiment). In our proposed set-up the charge-fluctuations are
described by a two-channel Anderson model for which we find that the charge
susceptibility is a regular quantity. Since the charge degrees of freedom are
directly treated, our predictions are not limited by an effective Kondo
temperature involved in mapping the two charges at the charge degeneracy point
into a spin degree of freedom.

In summary, we have made precise predictions for the capacitance lineshape of
two-channel quantum dots, together with a concrete proposal for testing the
same. As mentioned earlier, if the dot gate is set during the tuning stage in
a way of making $N_{d}$ odd then an $SU\!\left(  4\right)  $ single-channel
model results. The capacitance lineshape in such a situation, as for all
single-channel models, will be very different. In particular, a marked
non-zero skewness (making the plot asymmetric) should be seen when the
differential capacitance lineshape is plotted as a function of the excess
charge. If no assumptions are made, then while tuning the set-up the right
hand side of Eq.~(\ref{eqn: constraint}) can also turn out to be $N+2$. Such a
case is equivalent to the one analyzed above; one again has either a
two-channel or an $SU\!\left(  4\right)  $ model but this time for $N_{d}$ odd
or even respectively. Implications of this freedom and the detailed line-shape
in the $SU\!\left(  4\right)  $ case should be the subject of further study
and will be presented elsewhere. Exploring the possible use of the proposed
set up for transport measurements is an interesting open question, since that
might in the future allow access to other, more exotic, aspects of two-channel
Kondo physics.

\begin{acknowledgments}
We would like to acknowledge discussions with F.~B.~Anders, N.~Andrei,
S.~Florens, T.~Giamarchi, L.~I.~Glazman, K.~Le Hur, ~A.~Rosch, A.~Schiller,
P.~Simon and H. van der Zant. One of the authors (C.~J.~B.) was partially
supported by the MaNEP program of the Swiss National Science Foundation.
\end{acknowledgments}

\bibliographystyle{apsrev}
\bibliography{strings,kondo2004}

\end{document}